\documentclass[seceq,supplement]{ptptex} 

\title{
Current fluctuations in systems with diffusive dynamics, in and out of equilibrium
}
\author{
Vivien \textsc{Lecomte}$^{1,2,}$, Alberto \textsc{Imparato}$^{3,}$
and Fr\'ed\'eric \textsc{van Wijland}$^{4,}$
}

\inst{
$^1$D\'epartement de Physique de la Mati\`ere Condens\'ee, Universit\'e de Gen\`eve, 24 quai Ernest Ansermet, 1211 Gen\`eve 4, Switzerland\\
$^2$Laboratoire de Probabilit\'es et Mod\`eles Al\'eatoires (CNRS UMR 7599), Universit\'e Paris Diderot -- Paris 7, 2 place Jussieu, 75251 Paris cedex 05, France\\
$^3$Department of Physics and Astronomy, University of {A}arhus,
 Ny Munkegade, Building 1520, 8000 Aarhus, Denmark\\
$^4$Laboratoire Mati\`ere et Syst\`emes Complexes (CNRS UMR 7057), Universit\'e Paris Diderot -- Paris 7, 10 rue Alice Domon et L\'eonie Duquet, 75205 Paris cedex 13, France
}

\recdate{
October 2009}

\abst{
For diffusive systems that can be described by fluctuating hydrodynamics and by the Macroscopic Fluctuation Theory of Bertini {\it et al.}, the total current fluctuations display universal features when the system is closed and in equilibrium. When the system is taken out of equilibrium by a boundary-drive, current fluctuations, at least for a particular family of diffusive systems, display the same universal features as in equilibrium. To achieve this result, we exploit a mapping between the fluctuations in a boundary-driven nonequilibrium system and those in its equilibrium counterpart. Finally, we prove, for two well-studied processes, namely the Simple Symmetric Exclusion Process and the Kipnis-Marchioro-Presutti model for heat conduction, that the distribution of the current out of equilibrium can be deduced from the distribution in equilibrium. Thus, for these two microscopic models, the mapping between the out-of-equilibrium setting and the equilibrium one is exact.
}

\newcommand{\dd}{\text{d}}

\newcommand{\ee}{\text{e}}

\newcommand{\p}{\partial}

\newcommand{\eps}{\varepsilon}

\newcommand{\WW}{\mathbb{W}}
\usepackage{mathrsfs}
\newcommand{\F}{{\mathscr F}}

\begin{document}

\maketitle

\section{Why studying current fluctuations?}
In one of his 1905 papers, Einstein~\cite{einstein} establishes, for the motion of small spheres in suspension, a relationship between their diffusion constant and the fluid's viscosity, that is a relation between fluctuations in equilibrium and a response coefficient when the system is driven away from equilibrium by an infinitesimal force. This instance of a fluctuation-dissipation theorem is a particular case of the Green-Kubo relations~\cite{kubotodahashitsume} which state that, if $Q(t)=\int j(r,t')\dd t' \dd r$ denotes the total current associated to some locally conserved observable, then the variance of $Q$ in equilibrium teaches us directly about the transport properties of that very observable. For example, if $Q$ is the total particle current in a fluid with mean density $\rho$, then its variance $\sigma=\langle Q^2\rangle_c/t$ verifies~\cite{derridaJSM} 
\begin{equation}
\sigma=2D\rho^2 k_\text{\tiny B} T\kappa_T
\end{equation}
where $\kappa_T=\frac{1}{\rho}\frac{\p \rho}{\p P}$ is the isothermal compressibility and $D$ is the diffusion constant. That current fluctuations teach us about (slightly) nonequilibrium physics is an idea that can pushed forward: nonlinear response coefficients --the so-called Burnett coefficients-- can be related to higher cumulants of the current. The recent upsurge of interest for current fluctuations can be attributed to the numerical work of Evans, Morris  and Cohen~\cite{evanscohenmorriss} and to the mathematical breakthrough of Gallavotti and Cohen~\cite{gallavotticohen}, who showed the existence of an extended Einstein's relation applying to the entire distribution of the current instead of bearing on its first nontrivial moment. This relation takes the form of a particular symmetry property of the distribution of $Q$ and now goes by the name of fluctuation theorem. In fact, even for a system in equilibrium, current fluctuations tell us about how far the system has been wandering away from its typical realization. It was indeed recently realized that in order to cope with a given current fluctuation, the system may have to adopt a strongly heterogeneous configuration~\cite{bodineauderrida2, bodineauderrida3, appertderridalecomtevanwijland}.\\

The goal of the present paper is to identify what the generic properties of the current distribution are in systems whose dynamics is diffusive and that can be described by fluctuating hydrodynamics. We shall focus both on systems in equilibrium and on systems driven out of equilibrium by boundary constraints. How to exploit fluctuating hydrodynamics to obtain predictions regarding current fluctuations (and other physically relevant quantities) has been formalized by Bertini {\it et al.} into the Macroscopic Fluctuation Theory (see also~\cite{pilgram-et-al,jordan-et-al} in another context). The latter, albeit formulated in a physicist's language, will be at the basis of part of the results presented here. But the newest and strongest results of this work are concerned with a puzzling correspondence between the current distribution in and out of equilibrium in two well-studied microscopic models, the Simple Symmetric Exclusion Process (SSEP, a model for particle transport) and the Kipnis-Marchioro-Presutti (KMP) model for heat conduction.\\

Let us now formulate the results we have obtained. Let $Q(t)$ be the total (space and time integrated) current flowing through the system over a given time window $[0,t]$. Denoting by $j=\frac{Q(t)}{t}$, our interest goes to the distribution of $j$, which decays exponentially with the extensive variable $t$, as $t$ goes to infinity,
\begin{equation}
\text{Prob}\left\{\frac{Q(t)}{t}=j\right\}\sim \ee^{\pi(j) t}
\end{equation}
Alternatively, we shall focus on the generating function of $Q$,
\begin{equation}
Z(s,t)=\langle\ee^{-s Q(t)}\rangle
\end{equation}
which also plays the role of a dynamical partition function for space-time realizations of the process in which the current is constrained to adopt a given mean value fixed by the conjugate variable $s$. In this language, the generating function of the cumulants of the current, $\psi(s)$, given by 
\begin{equation}
\psi(s)=\lim_{t\to\infty}\frac{\ln\langle\ee^{-s Q}\rangle}{t}=\text{max}_j\{\pi(j)-s j\}
\end{equation}
can also be viewed as a dynamical free energy (the Legendre transform of the entropy-like function $\pi(j)$) whose physical content is related to the nature of the various dynamical phases able to convey a given current. As for its equilibrium thermodynamics counterpart, $\psi(s)$ is the quantity to investigate if one wishes to bring forth universal features. And indeed, we shall prove that $\psi$ takes a universal scaling form for systems in equilibrium. We shall further demonstrate that the same universal form holds for boundary-driven systems, with however some important restrictions on the phenomenological coefficients $D$ and $\sigma$. Finally, for both the SSEP and the KMP process we will show an exact equality between $\psi(s)$ calculated for the equilibrium system and that calculated out of equilibrium.\\

This paper is organized as follows. We begin in section
\ref{2exemples} by giving two examples of microscopic systems that can
be described by fluctuating hydrodynamics at a coarse-grained
scale. Then we explain in section \ref{saddle} how determining current
large deviations amounts to evaluating the saddle point contribution
of a path-integral. This technical step is then put to work in
equilibrium (section \ref{equilibrium}) and out of equilibrium
(section \ref{outofequilibrium}) where we recall for completeness some
of our previous results\cite{imparatolecomtevanwijland}. The new
results of this work are presented in sections \ref{exactSSEP} and
\ref{exactKMP}: there we explain how the distribution of the current
in the boundary driven SSEP or KMP can be deduced from its expression
in the absence of a drive, at the microscopic level. Physical conclusions and yet open questions
are gathered in section \ref{qouvertes}.

\section{Fluctuating hydrodynamics, two examples}\label{2exemples}
\subsection{The Simple Symmetric Exclusion Process (SSEP)}\label{exempleSSEP}
The Simple Symmetric Exclusion Process can be viewed as model for the transport of particles on a one-dimensional lattice in which each site can be occupied, at most, by one particle. Each particle hops randomly (with a unit rate) to either of its two nearest neighbors. The mutual exclusion constraint is the source of all interactions between particles. Denoting by $n_i(t')$ ($0\leq t'\leq t$) the local occupation number at site $i$ (a binary variable), we construct an occupation field $\rho(x,\tau)=n_i(t')$ which is assumed to possess smooth variations at the space and time scales $x=i/L$ and $\tau=t/L^2$. We refer to \cite{tailleurkurchanlecomte-2} for an explicit construction  of the required coarse-graining. It can be shown\cite{eyinklebowitzspeer1,eyinklebowitzspeer2} that the evolution of $\rho(x,\tau)$ is given by the following Langevin equation,
\begin{equation}
\p_\tau\rho=-\p_x j,\;\;j=-D\p_x\rho+\xi
\end{equation}
where the Gaussian noise $\xi$ has correlations $\langle\xi(x,\tau)\xi(x',\tau')\rangle=\frac{\sigma(\rho(x,\tau))}{L}\delta(x-x')\delta(\tau-\tau')$. The functions $D(\rho)$ and $\sigma(\rho)$ that appear in the expression of the local particle current $j$ are given by $D=1$ and $\sigma(\rho)=2\rho(1-\rho)$. 
\subsection{The Kipnis-Marchioro-Presutti (KMP) model}\label{exempleKMP}
We adopt the formulation of Giardin\`a {\it et al.}~\cite{giardinakurchanredig,giardinakurchanredigvafayi} that describes the Kipnis-Marchioro-Presutti model for heat conduction~\cite{kipnismarchioropresutti} in terms of a Langevin process. A collection of $L$ harmonic oscillators on a one-dimensional chain are subjected to the instantaneous thermal noise produced by their nearest neighbors. Let $x_j$ denote the position of oscillator $j$, whose evolution is given by 
\begin{equation}\label{defLangevinKMP}
\frac{\dd x_j}{\dd t}=-x_j+x_{j+1}\eta_{j,j+1}-x_{j-1}\eta_{j-1,j}
\end{equation}
where the It\^o convention is used and where the $\eta_{\ell,\ell+1}$'s are  Gaussian white noises with variance unity. The coupling of $x_j$ to its nearest neighbors arises through the local and fluctuating  temperatures $x_{j-1}^2$ and $x_{j+1}^2$ imposed by its two nearest neighbors. In this model, there is local conservation of the energy $\eps_j=\frac{x_j^2}{2}$. Assuming that the local energy field has smooth variations at the scales given by $x=j/L$ and $\tau=t'/L^2$ (with $0\leq x\leq 1$ and $0\leq\tau\leq t/L^2$, where $t$ is the macroscopic observation time), the theory of fluctuating hydrodynamics allows us to write that the local energy field $\rho(x,\tau)=\eps_j(t')$ evolves according to
\begin{equation}
\p_\tau\rho=-\p_x j,\;j=-D\p_x\rho+\xi
\end{equation}
where the Gaussian noise $\xi$ has variance $\langle\xi(x,\tau)\xi(x',\tau')\rangle=\frac{\sigma(\rho(x,\tau))}{L}\delta(x-x')\delta(\tau-\tau')$. For the KMP process of (\ref{defLangevinKMP}), the functions $D(\rho)$ and $\sigma(\rho)$ are given by
\begin{equation}
D(\rho)=1,\;\sigma(\rho)=4\rho^2
\end{equation}
We shall not prove this result here and we refer the reader to Bertini {\it et al.}~\cite{bertinidesolegabriellijonalasiniolandim-1,bertinidesolegabriellijonalasiniolandim-2,bertinidesolegabriellijonalasiniolandim-3,bertinidesolegabriellijonalasiniolandim-5,bertinidesolegabriellijonalasiniolandim-6,bertinigabriellilebowitz} and references therein for an introduction to the macroscopic fluctuation theory, and to \cite{tailleurkurchanlecomte-2} for a physicist's approach. 

\subsection{General framework}
We now summarize the hypotheses at the basis of fluctuating hydrodynamics. The relevant degrees of freedom, be they discrete (as in the SSEP) or continuous (as in KMP) are described at a coarse-grained level by a density field $\rho(x,\tau)$, in space units $x=i/L$ where the system size is unity and the running time is scaled by the typical diffusion time at the scale of the system's size, $\tau=t'/L^2$ ($0\leq t'\leq t$). At the scale given by the system size, fluctuations are asymptotically small, which accounts for the noise in the Langevin evolution equation (\ref{Langevinbasis})
\begin{equation}\label{Langevinbasis}
\p_\tau\rho=-\p_x j,\;\; j=-D\p_x\rho+\xi
\end{equation}
having a variance with a $1/L$ dependence,
\begin{equation}
\langle\xi(x,\tau)\xi(x',\tau')\rangle=\frac{\sigma(\rho(x,\tau))}{L}\delta(x-x')\delta(\tau-\tau')
\end{equation}
The weakness of the noise in the large system size limit is the key ingredient that makes our calculations possible, as we shall now present.
\section{A saddle point calculation}\label{saddle}
We start from the Langevin equation (\ref{Langevinbasis}) for the field $\rho(x,\tau)$ and from the expression of the total  time and space integrated current $Q(t)=L^2\int_0^{t/L^2}\dd\tau \int_0^1\dd x \; j(x,\tau)$, whose generating function we write in the form of a path integral based on the  Janssen-De~Dominicis~\cite{janssen-1,dedominicis} mapping:
\begin{equation}\label{partitionZ}
Z(s,t)=\langle\ee^{-s Q}\rangle=\int\mathcal D\bar{\rho}\mathcal D\rho\ee^{-L S[\bar{\rho},\rho]}
\end{equation}
where the action is expressed as
\begin{equation}
S=\int_{0}^{t/L^2}\!\!\!\dd\tau \int_{0}^{1}\dd x \;\left[\bar{\rho}\p_\tau\rho+D(\rho)\p_x\rho\p_x{\bar\rho}-\frac 12\sigma(\rho)(\p_x\bar{\rho}-sL)^2-(sL) D\p_x\rho\right]
\label{eq:action_rho_barrho}
\end{equation}
We denote by $\tilde{\rho}(x,\tau)=\bar{\rho}(x,\tau)-sL x$. As was pointed earlier~\cite{kurchan2, tailleurkurchanlecomte-2} the path integral in (\ref{partitionZ}) calls for a saddle point evaluation in the large system size limit $L\to\infty$. We denote by $\tilde{\rho}_c(x,\tau)$ and $\rho_c(x,\tau)$ the solutions to 
\begin{equation}\label{saddle1}\begin{split}
\frac{\delta S}{\delta\tilde{\rho}}=\p_\tau\rho-\p_x(D\p_x\rho)+\p_x(\sigma\p_x\tilde{\rho})=0\\
-\frac{\delta S}{\delta{\rho}}=\p_\tau\tilde{\rho}+\p_x(D\p_x\tilde{\rho})+\frac{\sigma'}{2}(\p_x\tilde{\rho})^2=0
\end{split}\end{equation}
Equations (\ref{saddle1}) must be complemented with the appropriate boundary conditions~\cite{tailleurkurchanlecomte-2}. To leading order in $L$ the partition function reads
\begin{equation}
Z(s,t)\sim\ee^{-L S[\tilde{\rho}_c,\rho_c]}
\end{equation}
We shall assume that the saddle point solution $(\tilde{\rho}_c,\rho_c)$ is stationary (this issue was discussed {\it e.g.} in \cite{bertinidesolegabriellijonalasiniolandim-5,bertinidesolegabriellijonalasiniolandim-6, bodineauderrida2}). This assumption, when not fulfilled, is signalled by instabilities that are interpreted as phase transitions~\cite{bertinidesolegabriellijonalasiniolandim-5,bertinidesolegabriellijonalasiniolandim-6, bodineauderrida2,bodineauderrida3,appertderridalecomtevanwijland,bodineauderridalecomtevanwijland}. Therefore, to leading order in the system size we have that
\begin{equation}\label{DBresult}
\psi(s)\Big|_\text{saddle}=\frac{\mu(sL)}{L},\;\mu(sL)=-\int_0^1\dd x\left[D(\rho_c)\p_x\rho_c\p_x\tilde{\rho}_c-\frac 12\sigma(\rho_c)(\p_x\tilde{\rho}_c)^2\right]
\end{equation}

Of course, as in any saddle point calculation, it is important to evaluate the leading corrections $\psi(s)\Big|_\text{fluct}$ to the asymptotic behavior given in (\ref{DBresult}). This is done by expanding the action $S$ around the saddle to quadratic order in the deviation from the saddle $\phi=\rho-\rho_c$ and $\bar{\phi}=\tilde{\rho}-\tilde{\rho}_c$,
\begin{equation}\label{actionquad}\begin{split}
S[\bar{\phi},\phi]=\int&\Big[\bar{\phi}\p_\tau\phi+D(\rho_c)\p_x\phi\p_x{\bar\phi}+D'(\rho_c)\p_x\tilde{\rho}_c\phi\p_x\phi+\frac 12 D''(\rho_c)\p_x\tilde{\rho}_c\p_x\rho_c \phi^2\\&+D'(\rho_c)\p_x\rho_c\p_x\bar{\phi}\,\phi-\frac 12 \sigma(\rho_c)(\p_x\bar{\phi})^2-\sigma'(\rho_c)\p_x\tilde{\rho}_c\phi\p_x\bar{\phi}-\frac 14 \sigma''(\rho_c) (\p_x\tilde{\rho}_c)\phi^2\Big]
\end{split}\end{equation}
and by integrating out the resulting quadratic form. Note that the latter step, which requires diagonalizing the quadratic form (\ref{actionquad}), may prove difficult when the coefficients of the quadratic form are space dependent, or, equivalently, if the saddle point solution is not homogeneous. In the next two sections, we implement the program we have just sketched in two distinct settings: for a closed equilibrium system and for an open system driven out of equilibrium by boundary constraints. 

\section{In equilibrium: closed systems with periodic boundary conditions}\label{equilibrium}
We first consider closed systems with periodic boundary conditions \cite{appertderridalecomtevanwijland}. The solution to the saddle point equations 
(\ref{saddle1}) is indeed rather simple to find, namely
\begin{equation}
\rho_c(x)=\rho,\;\tilde{\rho}_c(x)=-sL x
\end{equation}
where $\rho$ (with no argument) is the space averaged density. This leads to $\mu(\lambda)=\frac{1}{2}\sigma(\rho)\lambda^2$, which, with $\lambda=sL$, also reads $\psi(s)\Big|_\text{saddle}=L\frac 12 \sigma s^2$. Corrections to the saddle arising from integrating out the quadratic fluctuations around the optimal profile $\rho_c,\tilde{\rho}_c$ are not hard to evaluate, since the quadratic form (\ref{actionquad}) has constant coefficients. To do so we expand $\bar{\phi}$ and $\phi$ in Fourier modes indexed with wave vectors $q=2\pi n$, with $n\in\mathbb{Z}$, as imposed by the periodic boundary conditions. We find that the contribution of the determinant reads
\begin{equation}
\psi(s)\Big|_\text{fluct}=\frac{1}{2L^2}\sum_q\left[Dq^2-\sqrt{Dq^2\left(Dq^2-\frac
{\sigma\sigma''}{2D}(sL)^2\right)}\right]
\end{equation}
which we rewrite in the form
\begin{equation}
\psi(s)-\frac{\langle Q^2\rangle_c}{2t}s^2=\frac{D}{L^2}{\F}\left(\frac{\sigma\sigma''}{16D^2}(sL)^2\right)
\end{equation}
where $\F$ is a universal scaling function \cite{appertderridalecomtevanwijland}, a representation of which is given in terms of the Bernoulli numbers $B_{2n}$:
\begin{equation}
\F(x)=\sum_{k\geq 2}\frac{B_{2k-2}}{\Gamma(k)\Gamma(k+1)}(-2x)^k
\end{equation}
The scaling function $\F$ has a branch cut along the positive real axis when $x\geq \pi^2/2$. If the argument $x=\frac{\sigma\sigma''}{16D^2}(sL)^2$ of $\F$ hits the value $\pi^2/2$ upon varying $s$ this signals that the basic hypotheses underlying the saddle point calculation are not fulfilled, \emph{e.g.} that the stationary saddle point solution becomes unstable~\cite{bertinidesolegabriellijonalasiniolandim-5,bertinidesolegabriellijonalasiniolandim-6}. We refer the reader to Bodineau and Derrida~\cite{bodineauderrida2,bodineauderrida3} for an interpretation in terms of dynamic phase transitions.

At fixed value of $s$ and in the large system size limit $L\to\infty$, the limiting behavior of $\psi$ is given by
\begin{equation}
\frac 1L \psi(s)=\frac 12 \sigma s^2+\frac{\sqrt{2}}{3\pi}\sigma^{3/2}|s|^3 + o(|s{|}^3)
\end{equation}
 whose fourth derivative is singular at $s=0$. This was interpreted by Lebowitz and Spohn~\cite{lebowitzspohn}, in the particular case of the SSEP, in terms of the Burnett coefficients being infinite. This result is shown to apply irrespective of the explicit expression of $D$ and $\sigma$.

\section{Out of equilibrium: open boundary-driven systems}\label{outofequilibrium}
We now turn to an open system with the same bulk dynamics as that given by (\ref{Langevinbasis}), in contact at its boundaries with reservoirs that impose prescribed values for the field: $\rho(0,\tau)=\rho_0$ and $\rho(1,\tau)=\rho_1$. The saddle point equations (\ref{saddle1}) must now be solved bearing in mind these new boundary conditions. A stationary solution does exist, although it is now strongly space dependent. This should not be a surprise given that already at $s=0$, the optimal profile has a nonzero gradient allowing to bridge $\rho_0$ to $\rho_1$. In general, the explicit form of $\rho_c(x)$ and $\tilde{\rho}_c(x)$ is difficult to obtain. The function $\mu$ that appears in the rhs of (\ref{DBresult}), as calculated from plugging the solution (\ref{saddle1}) using the new boundary conditions into (\ref{DBresult}) is exactly the one that Bodineau and Derrida~\cite{bodineauderrida} initially found in their paper on the additivity principle. When $D(\rho)$ is a constant and $\sigma(\rho)$ is a quadratic function of $\rho$ then the analytics somewhat simplify  and it can be seen by direct calculation~\cite{imparatolecomtevanwijland} that, for $D=1$ and $\sigma(\rho)=c_1\rho+c_2\rho^2$, the saddle point contribution is given by
\begin{equation}\label{muomega}
\mu(\lambda)=\left\{
\begin{array}{ll}
-\frac{2}{c_2}(\operatorname{arcsinh}\sqrt{\omega})^2&\text{ for }\omega>0\\
+\frac{2}{c_2}(\arcsin\sqrt{-\omega})^2&\text{ for }\omega<0
\end{array}
\right.
\end{equation}
where $\omega(\lambda,\rho_0,\rho_1)$ is the auxiliary variable given by
\begin{equation}\label{def-omega-general}
\omega(\lambda,\rho_0,\rho_1)=\frac{c_2}{c_1^2}(1-\ee^{c_1\lambda/2})\left(c_1(\rho_1-\ee^{-c_1\lambda/2}\rho_0)-c_2(\ee^{-c_1\lambda/2}-1)\rho_0\rho_1\right)
\end{equation}
For the SSEP, $\sigma(\rho)=2\rho(1-\rho)$ and one recovers the known~\cite{derridadoucotroche,jordansukhorukovpilgram} result (the notation $z=\ee^{-\lambda}$ is used in the formula (2.14) of~\cite{derridadoucotroche}), namely
\begin{equation}\label{def-omega-SSEP}
\omega(\lambda,\rho_0,\rho_1)=(1-\ee^\lambda)(\ee^{-\lambda}\rho_0-\rho_1-(\ee^{-\lambda}-1)\rho_0\rho_1)
\end{equation}
For the KMP chain of coupled harmonic oscillators, the variable $\omega$ is now given by
\begin{equation}\label{def-omega-KMP}
\omega(\lambda,\rho_0,\rho_1)=\lambda(2(\rho_0-\rho_1)-4\lambda\rho_0\rho_1)
\end{equation}
The difficulty, at this stage, remains to diagonalize the quadratic form (\ref{actionquad}) given that its coefficients are space-dependent constants. The eigenmodes are not the standard plane waves anymore given that translation invariance does not hold. We have not been able to carry out this task in general, but we have found a way to bypass this technical step when $D$ is constant and $\sigma$ is a quadratic function of $\rho$. By introducing two auxiliary fields $\bar{\psi}$ and $\psi$ defined by
\begin{equation}\label{cov}
\phi=(\p_x\tilde{\rho}_c)^{-1}\psi+\p_x\rho_c\bar{\psi},\;\;\bar{\phi}=\p_x\tilde{\rho}_c\bar{\psi}
\end{equation}
which we substitute into (\ref{actionquad}), and after extensively using (\ref{saddle1}), we arrive at the following expression for $S$
\begin{equation}\label{actionquad2}\begin{split}
S=-\frac{\mu(sL)t}{L^2}+\int\dd x\dd\tau\left(\bar{\psi}\p_\tau\psi+D\p_x\bar{\psi}\p_x\psi -\mu(sL)(\p_x\bar{\psi})^2-\frac{\sigma''}{4}\psi^2\right) 
\end{split}\end{equation}
The local rotation of the fluctuation fields (\ref{cov}) has allowed to disentangle the space dependence and to find a set of variables in which translation invariance is recovered. The action (\ref{actionquad2}) exactly describes the quadratic fluctuations around the saddle in an open system {\it in equilibrium}, in which the parameter conjugate to the current is now $s'=\frac{\mu(sL)}L$. We diagonalize (\ref{actionquad2}) with the help of the Fourier modes $\{\sin qx\}_{q=n\pi}$, $n\in\mathbb{N}^*$ consistent with the field being fixed at the  $x=0$ and $x=1$ boundaries. The conclusion of this section is that for systems having a constant $D$ and a quadratic $\sigma$, we can actually determine the finite size corrections to the large deviation function and we find that
\begin{equation}
\psi(s)\Big|_\text{fluct}=\frac{D}{8L^2}{\F}\left(\frac{\sigma''}{2D^2}\mu(s L)\right)
\end{equation}
This is the very same function $\F$ that appears here for a boundary-driven open system as the one that was found when studying its closed equilibrium counterpart. We thus draw the partial conclusion that at least for a subclass of systems described by fluctuating hydrodynamics (those with constant $D$ and quadratic $\sigma$), the current distribution displays universal features, and these are the same as the ones observed in equilibrium. To reach this conclusion, we have resorted to a local mapping of the out-of-equilibrium system's fluctuations onto those of a corresponding equilibrium system.

\section{Exact mapping for the driven SSEP onto an equilibrium system}\label{exactSSEP}
Let us consider the evolution operator of the SSEP on a one-dimensional lattice with $L$ sites with injection rate at the left (resp. right)  boundary $\alpha$ (resp. $\delta)$ and annihilation rate at the left (resp. right) boundary $\gamma$ (resp. $\beta$). The hopping rate is set to 1. In the present section, and in the next, we find it more convenient to study the statistics of the total current flowing between the final site $L$ and the right reservoir, for which we denote the conjugate variable $\lambda$. It was shown explicitly in \cite{imparatolecomtevanwijland} that the formal replacement of $\lambda$ with $s(L+1)$ in the large deviation function allowed to pass from the current from the last site to the current flowing through the whole system. Thus we consider the evolution operator of the SSEP with the constraint that it has to carry a prescribed mean particle current (enforced by the Lagrange multiplier $\lambda$) between site $L$ and the rightmost reservoir. This evolution operator can be expressed in terms of the Pauli matrices $\sigma_j^x$, $\sigma_j^y$ and $\sigma_j^z$, and the raising and lowering operators $\sigma^\pm_j=\frac 12 (\sigma_j^x\pm i\sigma_j^y)$, whose algebra is given by
\begin{equation}
[\sigma^z,\sigma^\pm]=\pm 2\sigma^\pm,\;\;[\sigma^-,\sigma^+]=-\sigma^z
\end{equation}
It reads
\begin{equation}\label{defWL}\begin{split}
\mathbb{W}_L(\lambda)=&\frac 12 \sum_{j=1}^{L-1}\Big[\vec{\sigma}_j\cdot\vec{\sigma}_{j+1}-1\Big]+\alpha\left(\sigma_1^++\frac 12 \sigma_1^z-\frac 12\right)+\gamma\left(\sigma_1^--\frac 12 \sigma_1^z-\frac 12\right)\\
&+\delta\left(\ee^{\lambda}\sigma_L^++\frac 12 \sigma_L^z-\frac 12\right)+\beta\left(\ee^{-\lambda}\sigma_L^--\frac 12 \sigma_L^z-\frac 12\right)
\end{split}\end{equation} 
The parameter $\lambda$ is conjugate to the time-integrated current flowing from site $L$ to the right particle reservoir. Let us now consider a rotation of the spins $\vec{\sigma}_j=R\vec{s}_j$, where we write the $SO(3)$ matrix $R$ with a Cayley representation indexed by three parameters $x$, $y$ and $z$, namely $R =
  (I+A)(I-A)^{-1}$ with 
\begin{equation}
A=\left(
\begin{array}{lll}
 0 & -i z & y \\
 i z & 0 & -i x \\
 -y & i x & 0
\end{array}
\right)
\end{equation}
so that, explicitly,
\begin{equation}
R=\frac 1{1-x^2+y^2-z^2}
\left(
\begin{array}{lll}
 -x^2-y^2+z^2+1 & 2 i (x y-z) & 2 (y-x z) \\
 2 i (x y+z) & x^2+y^2+z^2+1 & -2 i (x-y z) \\
 -2 (y+x z) & 2 i (x+y z) & x^2-y^2-z^2+1
\end{array}
\right)
\end{equation}
We carry out the rotation of the spins in the evolution operator $\mathbb{W}_L(\lambda)$ which appears in (\ref{defWL}) and we search for a rotation $R$ that allows to interpret the resulting operator, when expressed in terms of the new variables $\vec{s}$, as an evolution operator for a driven and open SSEP with modified rates $\alpha'$, $\beta'$, $\gamma'$ and $\delta'$, and with a modified parameter $\lambda'$ constraining the particle current flowing out of the system. The resulting constraints read $y=-z$ and
\begin{align}
  \alpha' &=
  \frac{(1-z)\alpha-(x+z)\gamma}{1-x}
\\
  \gamma' &=
  \frac{(z-x)\alpha+(1+z)\gamma}{1-x}
\\
  \delta' &=
  \frac{(1-z) \big[1+x \ee^{\lambda}+z(1-\ee^{\lambda}) \big]\delta \:-\: (x+z) \big[x+\ee^{-\lambda}-z(1-\ee^{-\lambda}) \big] \beta  }{1-x^2}
\\
  \beta' &=
  \frac{(z-x) \big[x+ \ee^{\lambda}+z(1-\ee^{\lambda}) \big]\delta \:+\: (1+z) \big[1+x \ee^{-\lambda}-z(1-\ee^{-\lambda}) \big] \beta  }{1-x^2}
\end{align}
and the effective $\lambda'$ verifies
\begin{equation}
 \ee^{-\lambda'}=\frac
 {x+\ee^{-\lambda}+z(\ee^{-\lambda}-1)}
 {1+x\ee^{-\lambda}+z(\ee^{-\lambda}-1)}
\end{equation}
It is convenient to rewrite the above conditions in terms of the original densities $\rho_0=\frac{\alpha}{\alpha+\gamma}$ and $\rho_1=\frac{\delta}{\delta+\beta}$ and in terms of the auxiliary parameters $a=\frac{1}{\alpha+\gamma}$ and $b=\frac{1}{\delta+\beta}$. These now read, with obvious definitions of the primed quantities,
\begin{align}
 \rho'_0&=\frac
 {(1+x)\rho_0-x-z}
 {1-x}
 \label{eq:primedrhoa}
\\
 \rho'_1&=(x+\ee^{-\lambda}-z(1-\ee^{-\lambda}))\:\frac
 {\big[x+\ee^{\lambda}+z(1-\ee^{\lambda})\big]\rho_1 
  \: -x-z}
 {1-x^2}
 \label{eq:primedrhob}
\\
  a'&=a
\\
  b'&=b
\end{align}
At this stage, we have simply mapped our evolution operator describing the driven nonequilibrium SSEP with parameters $(\alpha,\beta,\gamma,\delta,\lambda)$ onto another driven SSEP with new parameters $(\alpha',\beta',\gamma',\delta',\lambda')$.\\

We now go one step further and we ask if there exists a rotation (that is a pair of variables $x$ and $z$) such that the primed process is in equilibrium, that is, such that the stationary densities $\rho_0'$ and $\rho_1'$ at the left and right reservoir are equal
\begin{equation}
\rho_0'=\rho_1'
\end{equation}
Such a condition can never be fulfilled at $\lambda=0$, but at $\lambda\neq 0$, a solution for $x$ and $z$ always exists. We have thus established that the nonequilibrium open and driven SSEP can be mapped onto an equilibrium open SSEP at arbitrary density.\\

It is interesting that we can exploit the freedom to choose the equilibrium density to which the original nonequilibrium process is mapped: density $\rho_0'=\rho_1'=\frac 12$ indeed plays a special role for the SSEP, since, at this very density, whatever the forcing strength $\lambda'$, the density profile remains flat at a value $1/2$ at the macroscopic level. This makes the computation of the current large deviation function, in equilibrium at density $\frac 12$ particularly easy. The condition $\rho_0'=\rho_1'=\frac 12$ leads to
\begin{equation}
\ee^{-\lambda'}=\left(\sqrt{\omega}+\sqrt{1+\omega}\right)^2,\;\;\omega=(1-\ee^\lambda)(\ee^{-\lambda}\rho_0-\rho_1-(\ee^{-\lambda}-1)\rho_0\rho_1)
\end{equation}
and hence 
\begin{equation}
 \psi_L(\lambda;\rho_0,\rho_1,a,b)=\psi_L\left(-2\ln(\sqrt{\omega}+\sqrt{1+\omega});\frac 12,\frac 12,a,b\right)
 \label{eq:mapping_psi}
\end{equation}
We know that in the large system size limit, $\psi_L\left(\lambda';\frac 12,\frac 12,a,b\right)=\frac{\lambda'^2}{4L}$ which immediately allows us to recover the result of \cite{derridadoucotroche},
\begin{equation}
\psi_L(\lambda;\rho_0,\rho_1,a,b)=\frac{1}{L}\left(\operatorname{arcsinh}\sqrt{\omega}\right)^2
\end{equation}
We have therefore shown that the cumulant generating function of the
current out of equilibrium can be inferred from that in equilibrium.

Moreover, an equality analogous to~(\ref{eq:mapping_psi}) holds for the full operator of evolution,
which implies that at fixed rates $a$ and $b$, the partition function $Z(s,t)=\langle\ee^{-s Q}\rangle$
depends on $\rho_0$, $\rho_1$ and $\lambda$ only through the variable $\omega$, for all time $t$ and  size $L$,
a result in the spirit of~\cite{derridadoucotroche,derridagerschenfeld}.
Last, the exact mapping of this section directly translates at
the level of the hydrodynamic fields $\rho$, $\tilde\rho$ of
section~\ref{saddle}.  Indeed,
following\cite{tailleurkurchanlecomte-2}, the
action~(\ref{eq:action_rho_barrho}) may be recovered from the
evolution operator through the correspondence $S^+=(1-\rho)
e^{\tilde\rho}$, $S^-=\rho e^{-\tilde\rho}$, $S^z=2\rho-1$. The
rotation corresponds to the change of fields
\begin{align}
  \rho'&=\frac{e^{-{\tilde\rho}} \left(x+y+(z-1) e^{{\tilde\rho}}\right) \left((\rho -1) e^{{\tilde\rho}} (x-y)-(z+1) \rho
   \right)}{1-x^2+y^2-z^2}
\\
  e^{\tilde\rho'}&=\frac{x+y+(z-1) e^{{\tilde\rho}}}{e^{{\tilde\rho}} (x-y)-z-1}
\end{align}
One checks by direct computation it leaves the bulk action invariant,
while the boundary conditions become $\rho'(0)=\rho'_0$,
$\rho'(1)=\rho'_1$ for $x,y,z$ solution of~(\ref{eq:primedrhoa}),
(\ref{eq:primedrhoa}). Choosing $\rho'_0=\rho'_1=\frac 12$, 
one checks that this change of fields becomes~(\ref{cov}) for the fluctuations around 
saddle.

\section{Exact mapping for the driven  KMP onto an equilibrium system}\label{exactKMP}
We consider a microscopic version of the KMP process described in subsection (\ref{exempleKMP}) where the leftmost (resp. rightmost) oscillator is coupled to a heat bath at temperature $T_0$ (resp. $T_1$) with an exchange rate $\gamma_0$ (resp. $\gamma_1$). The bulk dynamics given in (\ref{exempleKMP}) is unchanged but the contact with the heat baths is now described by
\begin{align}
\frac{\dd x_1}{\dd t}=&-\left(\frac 12+\gamma_0\right)x_1+x_2\eta_{1,2}-\sqrt{2\gamma_0 T_0}\xi_0\\
\frac{\dd x_L}{\dd t}=&-\left(\frac 12+\gamma_1\right)x_L+\sqrt{2\gamma_1 T_1}\xi_1-x_{L-1}\eta_{L-1,L}\\
\label{LangevinL}
\end{align}
where $\xi_0$ and $\xi_1$ are Gaussian white noises with unit variance.  The heat current flowing from oscillator $L$ to the bath on the right hand side is
\begin{equation}\label{courantL+1}
j_{L+1}=\gamma_1 x_L^2-\gamma_1T_1-\sqrt{2\gamma_1 T_1}x_L\xi_1
\end{equation}
where the It\^o convention is used. The Fokker-Planck evolution operator for the KMP process not only contains the contribution  given by~\cite{giardinakurchanredig, tailleurkurchanlecomte-2,giardinakurchanredigvafayi} that describes the unconstrained dynamics, but it also contains $\lambda$-dependent contributions that constrain the trajectories to carry a given mean current whose value is tuned by that of $\lambda$. We find that
\begin{equation}\label{defWLKMP}\begin{split}
\mathbb{W}_L(\lambda)=&\sum_{j=1}^{L-1}(\vec{K}_jJ\vec{K}_{j+1}+1/4)+\gamma_0\left[K_1^z+2T_0 K_1^-+\frac 12\right]\\
&+\gamma_1\left[K_L^z(1-2\lambda T_1)+2T_L K_L^-  +2\lambda(\lambda T_1-1)K_L^+ +\frac 12  \right]
\end{split}\end{equation}
where $K_j^+=x_j^2/2$, $K_j^-=\p_j^2/2$, $K_j^z=\p_j(x_j\cdot )-1/2$, $K_j^x= K_j^++K_j^-$, $K_j^y=-i(K_j^+-K_j^-)$. The $\lambda$-dependent terms in (\ref{defWLKMP}) can be found directly from a Kramers-Moyal expansion as deduced from the Langevin equation for $x_L$ (\ref{LangevinL}) and from the expression of the current (\ref{courantL+1}). These operators verify the so-called $SU(1,1)$ algebra relations
\begin{equation}\label{algebreK}
[K^z , K^{\pm} ] = \pm 2K^\pm,\;\;[K^− , K^+ ] = K^z
\end{equation}
The $SO(2,1)$ metric matrix $J$ has elements 
\begin{equation}
J=\left(\begin{array}{ccc}1&0&0\\0&1&0\\0&0&-1\end{array}\right)
\end{equation} 
For KMP, we search, as explained in \cite{lorente}, for a Cayley representation of $SO(2,1)$ isometries in the following way. We search for a matrix $A$ verifying $A^TJ+JA=0$. Such a matrix takes the general form
\begin{equation}
A=\left(\begin{array}{ccc}0&iz&y\\-iz&0&ix\\y&ix&0\end{array}\right)
\end{equation}
so that the matrix $R$ can now be cast in the form $R=(I+A)(I-A)^{-1}$, namely
\begin{equation}
R=\frac{1}{1+x^2-y^2-z^2}
\left(\begin{array}{ccc}
1 + x^2 + y^2 + z^2& 2 i(z + y x)& 2 (y - x z)\\
-2 iz + 2 iy x& 1 - x^2 - y^2 + z^2& -2 iz y + 2i x\\
2 (y + x z)& 2 i(z y + x)&  1 - x^2 + y^2 - z^2
\end{array}\right)
\end{equation}
To each matrix $R$ of $SO(2,1)$ one can associate a $SU(1,1)$ transformation that leaves the $K^\alpha$'s algebra (\ref{algebreK}) invariant. This allows us to now proceed along the lines of the reasoning carried out for the SSEP. We define $\vec{k}_j$ such that $\vec{K}_j=R\vec{k}_j$ and we ask whether the evolution operator (\ref{defWLKMP}), when expressed in terms of the new operators $\vec{k}_j$, can be interpreted as the evolution operator of an open and driven KMP process with modified bath and current-forcing parameters, $\gamma_0'$, $T_0'$, $\gamma_1'$, $T_1'$ and $\lambda'$. This is indeed the case provided $y=-x$ and 
\begin{align}
 T'_0&=\frac
 {T_0(1-z)-2x}
 {1+z}
\\
 T'_1&=\frac{(T_1(1-z+2x\lambda)-2x))(1-z+2x\lambda)}{1-z^2}
\\
  \gamma_0'&=\gamma_0
\\
  \gamma_1'&=\gamma_1
\end{align}
and the new $\lambda'$ is given by
\begin{equation}
 \lambda'=\frac{(1+z)\lambda}{1-z+2x \lambda}
\end{equation}
Note that the conditions $\gamma_0'=\gamma_0$ and $\gamma_1'=\gamma_1$ are analogous to the conditions $a'=a$ and $b'=b$ in the SSEP. There always exists a solution for $x$ such that the transformed dynamics describes current fluctuations in an equilibrium system, that is with $T_0'=T_1'$. The latter temperature is then parametrized by $z$. For each value of $(x,z)$, the combination $\omega=\lambda(T_0-T_1-\lambda T_0 T_1)$ is left invariant by passing to the primed variables, but we have not been able to exploit this fact to recover, by simple means, the result (\ref{muomega},\ref{def-omega-KMP}). Just as was the case for the SSEP, the $\lambda\to 0$ limit is singular and the mapping fails to hold in that limit. 
One checks however that (for instance imposing $T'_0=T'_1=1$), at fixed $\gamma_0$, $\gamma_1$ the spectrum of the operator depends on $T_0$, $T_1$ and $\lambda$ only through the variable $\omega$:
\begin{equation}
  \operatorname{Sp} \WW_L(\lambda;T_0,T_1,\gamma_0,\gamma_1) = \operatorname{Sp}  \WW_L\left(-2\ln(\sqrt{\omega}+\sqrt{1+\omega});1,1,\gamma_0,\gamma_1\right)
\end{equation}
a result similar to that of the SSEP (section~\ref{exactSSEP}), which seems to endow $\omega$ with a physical meaning yet to uncover.

\section{Open issues}\label{qouvertes}
It is well-known~\cite{spohn-2} that boundary driven systems develop
long-range correlations. It is thus, at first sight, rather puzzling
that a local mapping such as the one of section (\ref{exactSSEP}) or
(\ref{exactKMP}) allows to map a nonequilibrium situation onto an
equilibrium one. When constraining the dynamics to carry a prescribed
mean current imposed by a Lagrange multiplier $\lambda$, in the long
time limit, the physical states associated with a given value of
$\lambda$ do not display long range correlations. This can be seen by
combining the explicit evaluation of correlation functions, as done by
Bodineau {\it et al.}~\cite{bodineauderridalecomtevanwijland} with the
results of Imparato {\it et al.}~\cite{imparatolecomtevanwijland},
which gives a finite correlation length
$\ell(s)=\frac{D}{sL}(\sigma\sigma'')^{-\frac 12}$.  Since the
long-rangedness disappears at nonzero $\lambda$, it may be less
surprising that a local transformation does the trick.  In the limit
$\lambda\to 0$, the correlation length $\ell(s)$ becomes infinite
which restores the long-range correlations of the unbiased dynamics.
Such a simplification did not occur in
\cite{tailleurkurchanlecomte-1,tailleurkurchanlecomte-2} where density
large deviations were considered in the absence of a $\lambda$-drive,
which may account for the nonlocal transformations needed in that work
to map the nonequilibrium dynamics onto equilibrium dynamics.

We do not doubt that similar transformations can be found at the level
of fluctuating hydrodynamics (beyond quadratic fluctuations) for
systems belonging to the same family as the SSEP and KMP (with a
constant diffusion constant $D(\rho)$ and a quadratic noise variance
$\sigma(\rho)$). It would be interesting to see the explicit form of
the continuum analog of our (pseudo)rotations. But of course, a much
more interesting issue is whether our conclusions hold irrespective of
the particular form of $D(\rho)$ and $\sigma(\rho)$. But that's
another kettle of fish.

\section*{Acknowledgments}
We would like to thank C\'ecile Appert-Rolland, Thierry Bodineau,
Bernard Derrida, Julien Tailleur and Jorge Kurchan, with whom we have
several fruitful interactions in the course of this work.
V.L. was supported in part by the Swiss NSF under MaNEP and Division
II.

%

\end{document}